# Unraveling the varied nature and roles of defects in hybrid halide perovskites with time-resolved photoemission electron microscopy

Sofiia Kosar[1], Andrew J. Winchester[1], Tiarnan A. S. Doherty[2], Stuart Macpherson[2], Christopher E. Petoukhoff[1], Kyle Frohna[2], Miguel Anaya[2,3], Nicholas S. Chan[1], Julien Madéo[1], Michael K. L. Man[1], Samuel D. Stranks*[2,3], Keshav M. Dani*[1]

[1] *Femtosecond Spectroscopy Unit, Okinawa Institute of Science and Technology, 1919-1 Tancha, Onna-son, Okinawa, 904-0495, Japan.*

[2] *Cavendish Laboratory, University of Cambridge, JJ Thomson Avenue, Cambridge CB3 0HE, United Kingdom.*

[3] *Department of Chemical Engineering and Biotechnology, University of Cambridge, Philippa Fawcett Drive, Cambridge CB3 0AS, United Kingdom.*

*kmdani@oist.jp, *sds65@cam.ac.uk

**With rapidly growing photoconversion efficiencies, hybrid perovskite solar cells have emerged as promising contenders for next generation, low-cost photovoltaic technologies. Yet, the presence of nanoscale defect clusters, that form during the fabrication process, remains critical to overall device operation, including efficiency and long-term stability. To successfully deploy hybrid perovskites, we must understand the nature of the different types of defects, assess their potentially varied roles in device performance, and understand how they respond to passivation strategies. Here, by correlating photoemission and synchrotron-based scanning probe X-ray microscopies, we unveil three different types of defect clusters in state-of-the-art triple cation mixed halide perovskite thin films. Incorporating ultrafast time-resolution into our photoemission measurements, we show that defect clusters originating at grain boundaries are the most detrimental for photocarrier trapping, while lead iodide defect clusters are relatively benign. Hexagonal polytype defect clusters are only mildly detrimental individually, but can have a significant impact overall if abundant in occurrence. We also show that passivating defects with oxygen in the presence of light, a previously used approach to improve efficiency, has a varied impact on the different types of defects. Even with just mild oxygen treatment, the grain boundary defects are completely healed, while the lead iodide defects begin to show signs of chemical alteration. Our findings highlight the need for multi-pronged strategies tailored to selectively address the detrimental impact of the different defect types in hybrid perovskite solar cells.**

An inherent downside to low-cost solution processed hybrid perovskite photovoltaic thin films is the generation of a variety of defects that impact device operation. These defects can include atomic vacancies and interstitials,[1,2] unreacted precipitates from the solution processing steps,[3,4] as well as alternative crystalline phases,[5,6] which limit device performance.[7-10] Designing successful strategies to mitigate these defects strongly depends on our fundamental understanding of their identity and role in performance. This requires a cumulative approach that visualizes defects with high spatial resolution, as well as accesses their exact roles in the operation of perovskite photovoltaic devices. Recently, we used photoemission electron microscopy (PEEM) and scanning electron analytical techniques to visualize nanoscale defect clusters and obtain some understanding relating to the structure and composition of the surrounding grains.[8] Further, time-resolved PEEM (TR-PEEM) techniques shed light on the general role of defects in the ultrafast charge carrier trapping process. Correlating local defect locations with nanoscale changes in structure upon light exposure has also revealed their role in long-term stability.[10] However, considering the variety of defects that can occur, the impact of specific types of defects on film performance remains largely a mystery. Moreover, while general strategies have been pursued thus far to address unwanted defects, uncovering the specific impact of these strategies on the different defect types is key to designing more-effective, targeted approaches.

Here, using time- and spectrally-resolved PEEM on state-of-the-art triple cation mixed halide perovskite films, we observe three different types of nanoscale surface defect clusters, and find a surprisingly varied effect on device efficiency and response to controlled treatment with light and oxygen. Defect clusters appearing exclusively at grain boundaries are small, on the order of a few tens of nanometres in size. Via diffusion assisted hole-trapping, they impact the performance of a region nearly ten times their size. We also observe lead iodide defect clusters that are relatively benign to device performance as they do not participate in the trapping of carriers generated in the perovskite films. Lastly, polytype defect clusters can be as large as few hundreds of nanometres, and exhibit altered crystalline phases to the pristine perovskite. Located nearly an electron-volt below the Fermi level, they trap holes in a region about twice their size, indicating a reduced zone of impact. We show that just mild doses of visible light and dry air completely eliminate the grain boundary defects while polytypes remain largely unchanged. The sites of lead iodide defect clusters show signs of chemical alteration on exposure, which could be an important factor in long-term stability of devices.[10] The presence of multiple defect types, and their varied roles in device efficiency and stability, suggests the need for targeted approaches to improve the performance of perovskite devices.

**Results and discussion**

For our study, we prepared triple cation mixed halide perovskite thin films with chemical composition of $FA_{0.78}MA_{0.17}Cs_{0.05}Pb(I_{0.83}Br_{0.17})_3$ by spin-coating the precursor solution onto glass substrates coated with indium tin oxide (see Experimental methods for details of fabrication process). Our samples showed emission peaks at ~ 772 nm, thus yielding a band gap of ~ 1.6 eV (Fig. S1, SI). We also deposited fiducial gold markers on the sample surface to help accurately superimpose images taken with different techniques[11] and to establish the Fermi level of samples in our photoemission experiments. We then transferred our samples from the $N_2$-filled glovebox to our PEEM utilizing a hermetically sealed vacuum suitcase to avoid exposure to ambient air. We took care to mitigate the ultraviolet light exposure conditions to ensure no sample damage was induced during PEEM measurements (see Methods section of our previous work[8]). In this study, we measured multiple samples and observed qualitatively similar results. As previously reported,[8] a PEEM image acquired when probing the surface of perovskite samples with 4.65 eV probe photons (Fig. 1a, see Experimental methods for imaging details) showed nanoscale clusters of occupied surface defect states (Fig. 1b, green contrast). Then, utilizing 6.2 eV probe photons, we measured the onset of the valence band and thereby imaged the surface morphology exhibiting the typical grain structure (Fig. 1b, grey contrast, see Experimental methods for imaging details). By overlaying the two images, we obtained a direct visual of the location of the defect clusters relative to the surface morphology (Fig. 1b). We immediately saw hints of the presence of multiple types of defects. Smaller defect clusters of few tens of nanometres were located predominantly at the grain boundaries, while larger clusters corresponded to sub-grain features or even occupied entire grains.

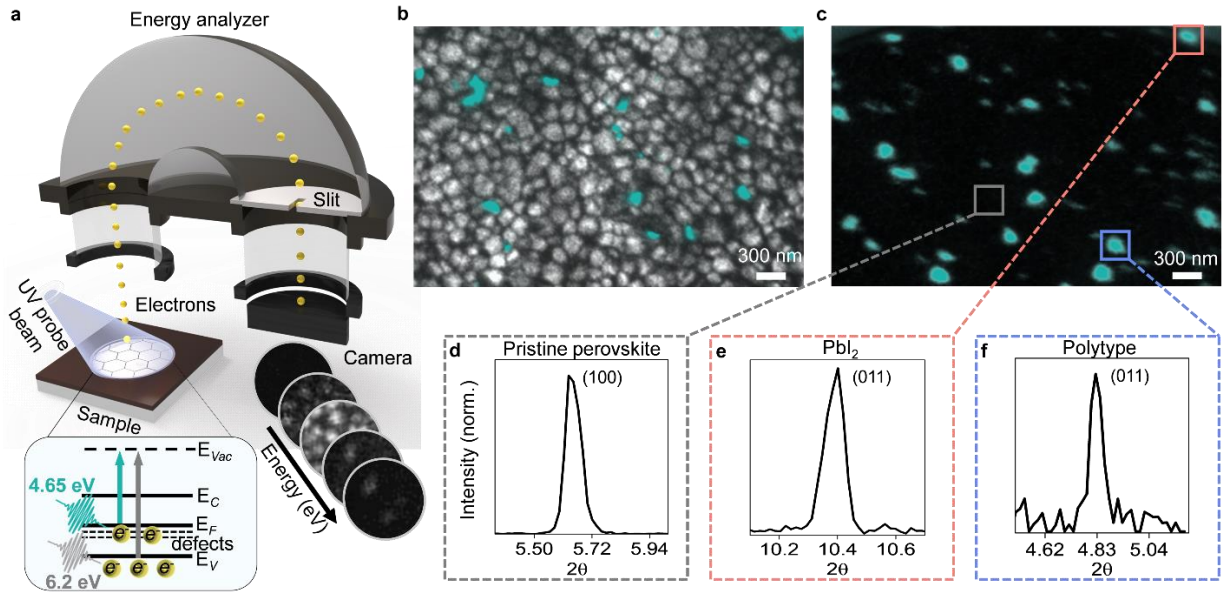

**Fig. 1 Presence of multiple types of defects on nanoscale.** (a) Schematic of PEEM experimental setup: UV probe pulses cause photoemission of electrons from sample surface; electrons are then accelerated through hemispherical energy analyzer, dispersed in energy, and imaged on CCD camera for each energy. Inset shows energy level diagram of hybrid perovskite: arrows indicate transitions from defect states (*defects*) below the Fermi level ($E_F$), valence band states ($E_V$), conduction band states ($E_C$) to vacuum states ($E_{Vac}$) upon probing of sample with photon energies of 4.65 eV (green arrow) and 6.2 eV (grey arrow). (b) Overlaid PEEM image of surface morphology obtained by illuminating the sample with 6.2 eV photons (grey contrast) and image of defect clusters acquired with 4.65 eV photons (green contrast). (c) PEEM image taken with 4.65 eV probe photons, overlaid with colored boxes showing nXRD regions of interest. (d) Local diffraction pattern extracted from the pristine region marked grey in (c) showing the (100) peak of a cubic perovskite. (e) Local diffraction pattern extracted from the defect cluster marked with a red box in (c) showing the (011) peak of $PbI_2$. (f) Local diffraction pattern extracted from the defect cluster marked with a blue box in (c) showing the (011) peak of a 6H hexagonal perovskite polytype.

To distinguish these defect clusters, we correlated synchrotron-based nano x-ray diffraction (nXRD) with PEEM measurements on the same region of the sample. The spatially averaged nXRD pattern extracted from the PEEM mapped region (Fig. 1c) can be indexed to a cubic perovskite structure with a lattice parameter of 6.3 Å, with trace amounts of $PbI_2$ also present but no other phase impurity peaks visible in the averaged pattern (Fig. S2, SI). Extracting local diffraction patterns from pristine regions away from the defect clusters revealed cubic perovskite structure (Fig. 1d). However, local diffraction patterns extracted from the larger PEEM defect clusters showed that some of these sites corresponded to $PbI_2$ (Fig. 1e). Inclusions of $PbI_2$ are known to form due to excess unreacted $PbI_2$ from the precursor solution.[12] Other large PEEM defect clusters exhibited the diffraction pattern of a hexagonal perovskite polytype (Fig. 1f). Hexagonal polytypes are intermediate products that form during crystallization of the perovskite thin film from precursor solution[13,14] and are susceptible to the formation of defects.[15] The grain boundary defects could not be unambiguously identified in nXRD due to their small size below the resolution of the nXRD measurements (~ 100 nm).

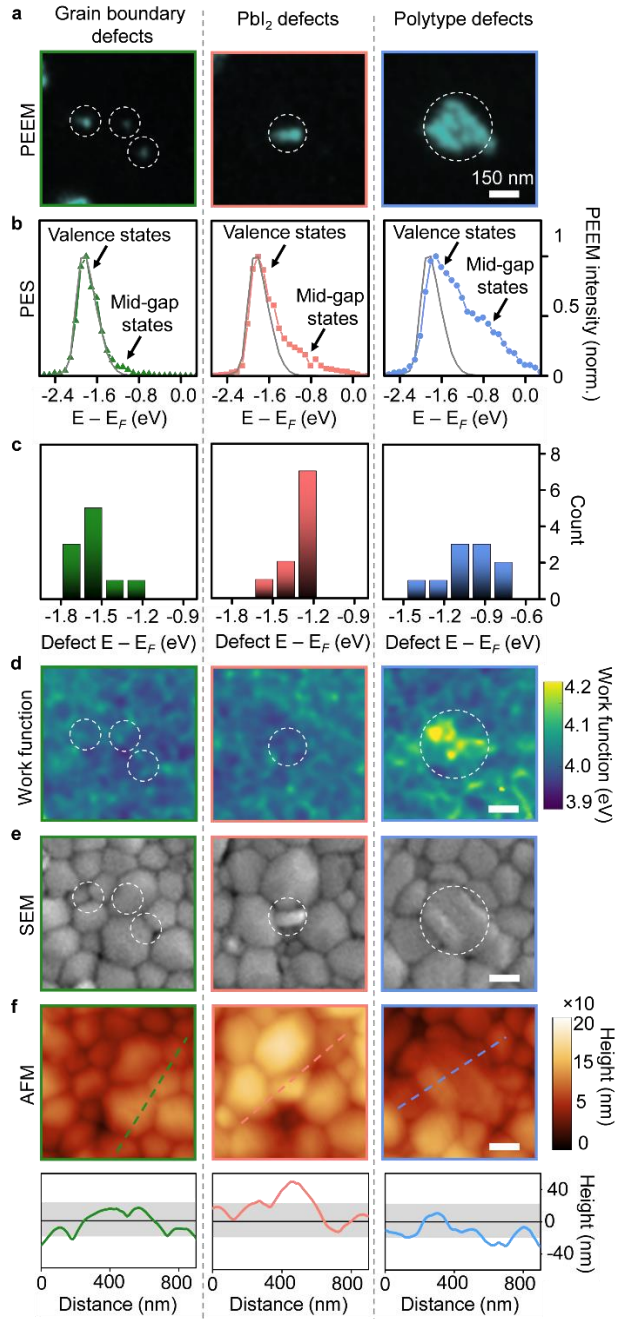

**Fig. 2 Nature of multiple defect types.** (a) High resolution PEEM images for three types of selected defect clusters, as labelled. (b) Representative photoelectron spectra for three types of defect clusters exhibiting both contributions from valence and mid-gap states. Grey line shows signal from a representative pristine region with main contribution from valance states. (c) Estimated peak energies of the mid-gap defect states for the three types of defects plotted based on fitting of photoemission spectra (Fig. S3, SI). (d) Work function maps for the same regions as (a). (e) SEM images for same regions as (a, d). (f) AFM images for the same regions as (a, d, e) with line profiles through the defect clusters indicating the raised height only for the PbI$_2$ defect clusters. Solid black line indicates mean height for the sample, with grey shaded area denoting the root-mean-square bounds.

To further distinguish these large defect clusters, as well as the small grain boundary defects, we used spectroscopic PEEM to obtain a nanoscale map of the photoemission spectrum as well as the local work function (see Experimental methods). We also utilized scanning electron (SEM) and atomic force (AFM) microscopies to aid in the interpretation of our results. Photoemission spectra from the small grain boundary defect clusters (Fig. 2a) exhibited a low mid-gap defect density (Fig. 2b) that was located closer to the valence band in energy (Fig. 2c, Fig. S3, SI). Given their size and location at the grain boundaries (Fig. 2e and f), we were not able to resolve obvious work function changes (Fig. 2d) at sites of grain boundary defect clusters. Photoemission spectra from $PbI_2$ defect clusters (Fig. 2a) exhibited a larger mid-gap defect density (Fig. 2b) and were located closer to the Fermi level in energy (Fig. 2c). In our work function maps, the $PbI_2$ defect clusters showed only a small increase in work function (by ~ 40 meV) compared to surrounding regions (Fig. 2d, Fig. S4, SI). Additionally, AFM mapping revealed that these defective $PbI_2$ inclusions were elevated in height above the perovskite grains (Fig. 2f), suggesting they are precipitates from the precursor solution. These can also be seen as bright high-contrast features in an SEM image (Fig. 2e).[4,16,17] Photoemission spectra from the polytype defect clusters (Fig. 2a) showed a defect density in the mid-gap region centered ~ 1eV below the Fermi level (Fig. 2b and c, Fig. S3, SI). The defective polytype grains (Fig. 2e and f), also showed a clear increase in work function (~ 80 meV) over surrounding regions (Fig. 2d, Fig. S4, SI), as expected for polytypes.[18]

The presence of multiple types of defects immediately suggests that tailored approaches may be necessary to passivate them. However, any meaningful strategy first requires knowledge about the specific role played by each defect type in device performance. For this, we turn to time-resolved photoemission electron microscopy[8,19-24] to understand their role in non-radiative recombination.[8] Accordingly, in the PEEM, we photoexcited carriers just at the band edge of the pristine perovskite material in our samples with near-infrared pump pulses (Fig. 3a), and imaged the change in photoemission from the mid-gap defect states with time-delayed 4.65 eV probe pulses (Fig. 3a, inset). Such images allowed us to analyze the role in charge carrier trapping for each defect cluster. For the pristine areas without defects, we did not observe changes in photoemission intensity (Fig. 3b), thus indicating the absence of trapping. Surprisingly, the three different defect types showed strikingly different behavior (Fig. 3b, Fig. S5, SI). For the $PbI_2$ defects, we did not observe substantial changes in photoemission intensity after photoexcitation (Fig. 3b and c), indicating that these defects were relatively benign from the perspective of trapping charges generated in the pristine perovskite material. We expect that such processes could be inhibited in the $PbI_2$ defects due to the morphological height barrier with the surrounding perovskite grains. Nonetheless, we note that one does observe localized photoluminescence losses at the $PbI_2$ defect clusters, due to the presence of non-perovskite material (Fig. 3d).

In contrast to the $PbI_2$ defect clusters, both the grain boundary and polytype defect clusters participated in carrier trapping, as seen by a decrease in photoemission intensity at the corresponding sites after photoexcitation (Fig. 3b). However, there were important differences between the two. Grain boundary defect clusters exhibited a large reduction in the TR-PEEM intensity after photoexcitation indicating a significant cross-section for photo-hole capture.[8] We used double exponential decays to characterize their dynamics (Fig. 3c), and revealed time constants of $\tau_1$ ~ 6 ps and $\tau_2$ ~ 290 ps, suggesting a local, fast trapping process, followed by a slower diffusion assisted trapping process – in good agreement with previous reports.[8] Photoluminescence maps around the grain boundary defect clusters showed the impact of this diffusion assisted trapping – the photoluminescence intensity was depleted over an area nearly ten times larger than the defect cluster size (Fig. 3d). In contrast, polytype defect clusters showed a much smaller

TR-PEEM response (Fig. 3b and c) with only a fast-exponential decay ($\tau_1 \sim 3.7$ ps) observable within our measurement window. This indicates that diffusion-assisted processes, if present, are weaker and slower. Consistent with this, local photoluminescence maps around the polytype defect clusters showed losses that extended only about twice the size of the polytype defect clusters (Fig. 3d). Fig. S6, SI shows additional regions with photoluminescence maps around defect clusters of each type that show the same qualitative behavior. We note that in samples that show a large abundance of polytype phases, despite the individual defect clusters being less detrimental than grain boundary types, they can have a comparable or larger impact on the overall film performance (Fig. S5, SI).

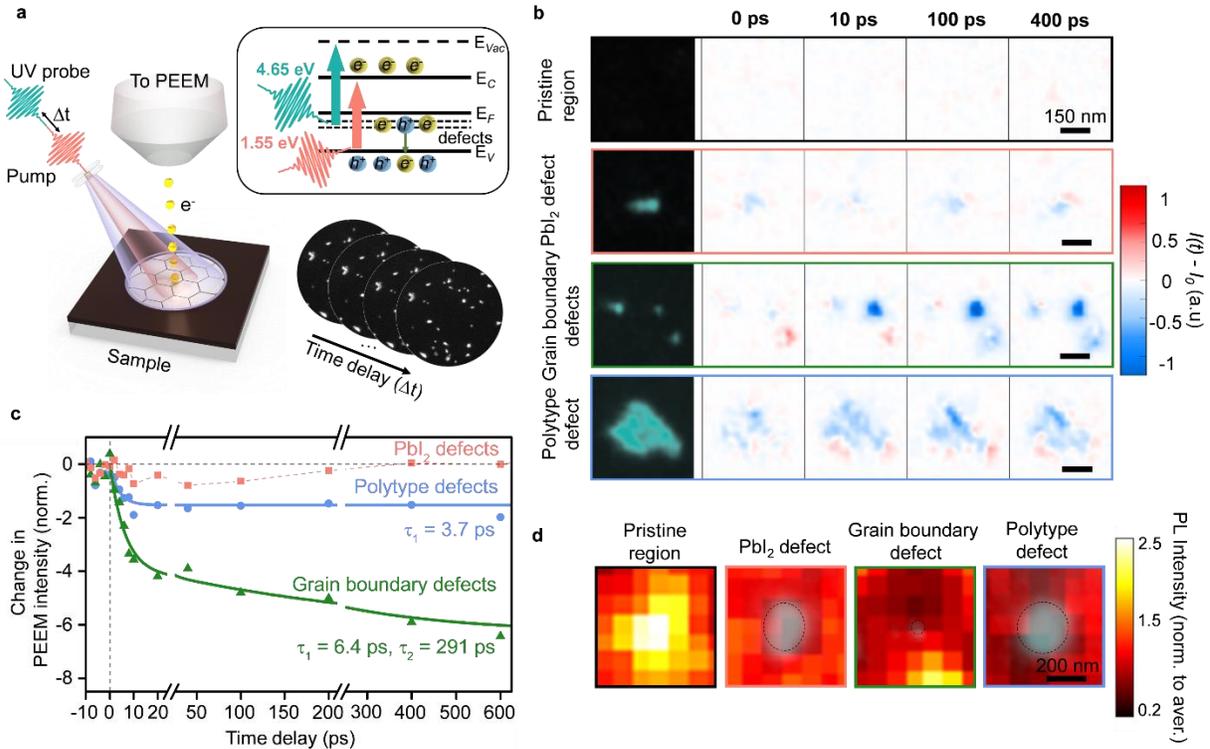

**Fig. 3 Role of different types of defect clusters in performance losses.** (a) Schematic of TR-PEEM experiment: sample is photoexcited with a pump pulse and probed with a time-delayed UV pulse. Photoemission signal is recorded as a sequence of images. Inset shows energy level diagram for perovskite sample: red arrow shows transition from valence band ($E_V$) to conduction band ($E_C$) upon photoexciting sample with pump photons; green arrow shows transitions from mid-gap defect levels (*defects*) below the Fermi level ($E_F$) to vacuum levels ($E_{Vac}$) upon probing the sample with 4.65 eV photons in PEEM. (b) Static PEEM images (left column) of a pristine region with no defects followed by regions with the three types of defect clusters, as labelled. Remaining columns denote the TR-PEEM intensity changes as ($I(t) - I_0$), in those regions for four different pump-probe delays: 0 ps, 10 ps, 100 ps and 400 ps. (c) TR-PEEM traces, i.e. percentage change in PEEM intensity after photoexcitation ($[I(t) - I_0]/I_0 \times 100$) plotted versus pump-probe time-delay for the three types of defect clusters. Solid lines represent exponential decay fits. (d) Photoluminescence maps overlaid with PEEM images (green contrast) for a pristine region, and regions exhibiting the three different types of defect clusters. PEEM intensity here plotted on logarithmic scale.

Over the past few years, to mitigate their detrimental impact, passivation strategies have been developed to address defects in general. Yet, the presence of different types of defects and their varied roles in photocarrier trapping suggests that it is important to consider how each defect type may respond to a passivation strategy. To understand this, we employed an approach previously used to enhance optoelectronic properties of perovskite thin films – treatment with simple small molecules, such as oxygen in the presence of light.[25-29] To differentiate the response of the different types of defects, we introduced only small, controlled doses of dry air and visible light (1 mbar dry air pressure and less than 1 sun illumination for 1 hour), which have been sufficient to begin showing signs of improved photoluminescence.[27,28] Strikingly, we observed complete removal of the most detrimental grain boundary defect clusters at these small doses, as seen from TR-PEEM measurements; no evidence of the previously seen photohole trapping at their sites remained (Fig. 4a and b). In contrast, defect densities (Fig. S7, SI) and trapping by polytype and $PbI_2$ defect clusters were largely unchanged (Fig. 4c and d, Fig. S8, SI). Interestingly, for $PbI_2$ defect clusters, we detected changes in work function, suggesting chemical alterations at these sites (Fig. 4d, Fig. S7, SI). These alterations are likely initial indicators of processes that occur with higher dosages of oxygen.[10]

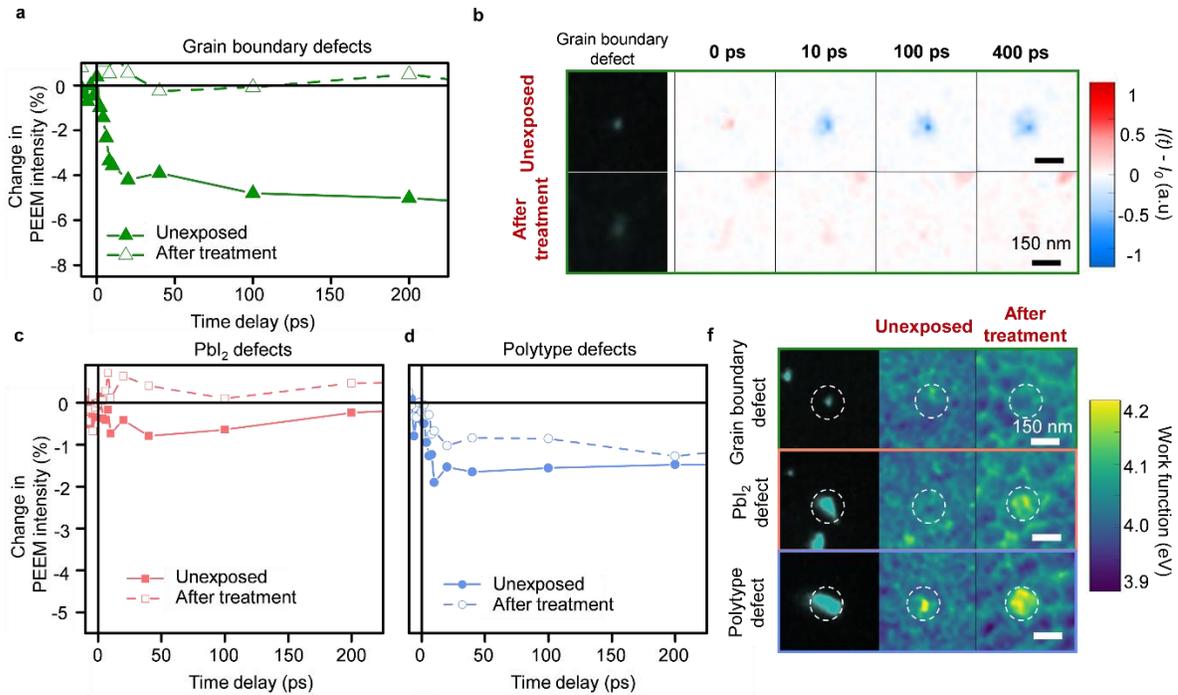

**Fig. 4 Response of different types of defect clusters to treatment with visible light and mild dry air environment.** (a) TR-PEEM decay curves averaged for multiple grain boundary defect clusters plotted as ($[I(t) - I_0]/I_0 \times 100$) for unexposed case and after treatment with 1 mbar dry air pressure upon illumination, as labelled. (b) PEEM images of the representative grain boundary defect cluster for unexposed case and after treatment with visible light and dry air. Remaining columns show the TR-PEEM signal at different time-delays for the corresponding regions. (c) TR-PEEM decay curves averaged for multiple defect clusters of the same type before and after the treatment with dry air upon illumination for $PbI_2$ defect clusters. (d) TR-PEEM decay curves before and after the treatment for polytype defect clusters. (f) PEEM images for selected defect clusters as labelled (left column) and local work function maps (right columns) for unexposed case and after treatment with dry air upon illumination.

## Conclusions

Our observation of multiple types of defects, which originate from different nanoscale compositions and play a surprisingly varied role in charge trapping, helps reconcile several previous observations and adds deeper insight into performance of hybrid perovskite films. Previous reports on high concentration of iodine species at the grain boundaries,[30] hole trapping by deep level iodine interstitials[1,31,32] and facilitated iodine ion migration at the grain boundaries[33] suggest that grain boundary defect clusters contain a high density of interstitial iodine. The immediate healing of these highly detrimental defects by exposure to light and a mild dry air atmosphere, suggests a potential approach to remove one important channel of non-radiative recombination. Furthermore, several literature reports highlight passivation of grain boundary defects after the addition of moderate amount of PbI$_2$ resulting in an increase of open circuit voltage,[16] enhanced carrier lifetimes,[4,33] and suppressed current-voltage hysteresis.[17] The relatively benign nature of PbI$_2$ defects versus the detrimental grain boundary defects suggests why these strategies appear to be successful. Previous reports show that, over time, material decomposition can lead to the formation of increased polytype phases.[34] Our work shows that in abundance these defective polytype phases become the dominant mechanism for photocarrier trapping and in limiting device efficiency. Thus, the development of targeted strategies to mitigate the formation or detrimental impact of these polytype defects, as well as other defect types, will benefit the long-term performance of photovoltaic devices.

## Experimental methods

### Perovskite preparation

Samples for the measurements have been prepared in Cambridge in a N$_2$-filled glovebox and shipped to Okinawa Institute of Science and Technology in sealed packaging to guarantee no exposure to ambient atmosphere or light.

Perovskite samples with chemical composition of (FA$_{0.78}$MA$_{0.17}$Cs$_{0.05}$)Pb(I$_{0.83}$Br$_{0.17}$)$_3$ have been spin-coated from precursor solutions containing (1 M) formamidinium iodide, (0.2 M) methylammonium bromide, (1.1 M) lead iodide and (0.22 M) lead bromide dissolved in 4:1 (v:v) anhydrous dimethylformamide (DMF) / dimethyl sulfoxide (DMSO). Cesium iodide has been first dissolved in DMSO (1.5 M), and then added to the precursor solution.

ITO substrates (1 cm × 1 cm in size) were sonicated in acetone and isopropanol (10 minutes per step) and then treated for 1 minute in an oxygen plasma cleaner. For nXRD/PEEM correlations, SiN windows (Norcada; NX7100c) were treated for 1 minute in an oxygen plasma cleaner. Precursor solution with triple cation mixed halide perovskite has been spin-coated on the substrates in a two-step manner: 2,000 rpm for 10 seconds and 4,000 rpm for 35 seconds. After 20 seconds into the second step, 100 μl of chlorobenzene was dropped on the substrate. To finish crystallization films were annealed at 100 °C for 1 hour to achieve average grain size on the order of 250 nm – 500 nm, and film thickness of ~ 500 nm.

We used gold fiducial markers for position reference for all the samples.[8,11] Gold fiducial markers have been prepared as a suspension of Au nanoplatelets in chlorobenzene, and spin-coated on the ready perovskite films at 1,000 rpm for 20 seconds. Gold markers varied in size between few micrometers to few tens of micrometers, and had various shapes, which made it easy to use them as position reference for measurements and overlay analysis. We note that Au markers do not affect our measurements.

### PEEM and TR-PEEM experimental setup

Photoemission electron microscope (SPELEEM III, Elmitec GmbH) with hemispherical energy analyzer was coupled with ultrafast laser source (FemtoLasers XL:650) that delivers pulses of 45 fs, 650 nJ at 800 nm, and operates at 4 MHz. We used custom-built optical setup to generate 3rd and 4th harmonics from fundamental laser output (4.65 eV and 6.2 eV photons, respectively) via sum frequency generation. The layout of experiment has been reported in Methods section of our previous work.[8]

**PEEM imaging**

Perovskite samples have been mounted on special cartridges for PEEM imaging inside of the $N_2$-filled glovebox. Cartridges with samples have been transferred to the microscope inside of a hermetically sealed transfer suitcase to avoid exposure to ambient air. The base pressure in the main imaging chamber of the microscope was kept at $10^{-10}$ - $10^{-11}$ torr.

High resolution PEEM images of the defect clusters were acquired at 10 µm field of view using 4.65 eV probe photons, and contrast aperture. PEEM images of surface morphology have been acquired at 10 µm field of view with 6.2 eV probe photons, contrast aperture and energy analyzer slit set to 125 meV. TR-PEEM images have been acquired at 10 µm field of view without contrast aperture. Energy-resolved images for photoelectron spectra have been acquired with 6.2 eV photons, without contrast aperture, with energy analyzer slit set to 125 meV, thus providing approximately 125 meV energy resolution.

The probe photon fluence was kept below 100 nJ/cm$^2$ for 4.65 eV photons, and below 10 nJ/cm$^2$ for 6.2 eV photons for all measurements. Probe fluences were kept low during the experiments to avoid sample damage and space charge that usually causes spectral broadening in PEEM experiments and reduces PEEM image resolution. The fluence of 1.55 eV pump photons for TR-PEEM measurements was below 100 µJ/cm$^2$.

PEEM images of defects have been taken with typical imaging conditions of 10 – 15 seconds exposure per frame, and averaged between 25 – 35 frames. Images of surface morphology have been acquired with 60 – 80 seconds exposure per frame, and averaged between 15 – 20 frames. Typical imaging conditions for TR-PEEM images were 1 second exposure and 64 averages per frame (1 delay step). Scans have been repeated 4 times, and final images are result of 4 averages. Energy-resolved images for photoelectron spectra were obtained with 20 seconds exposure, and 4 averages per image.

Each PEEM image has been processed with flat field method to remove inhomogeneity originating from image-acquiring parts of the instrument. Multiple frames of one image have been drift-corrected before averaging.

**Nano x-ray diffraction measurements**

Measurements were performed on the i14 Hard X-ray Nanoprobe beamline at the Diamond Light Source, Didcot, UK using a 20keV monochromated X-ray beam.[35] The diffracted X-rays were collected in a transmission geometry with an Excalibur 3M detector consisting of 3 Medipix 2048 x 512 pixel arrays. For all measurements involving nXRD/PEEM correlations, samples were deposited on X-Ray transparent, SiN windows from Norcada (product number NX7100C). The step size for the measurements was 100 nm and the dwell time was 0.75 s per point. All samples were stored in a nitrogen filled glovebox before measurement, and held under a nitrogen flow during measurements. 2D diffraction patterns were calibrated with a $CeO_2$ reference standard and radially integrated in the Data Analysis Workbence (DAWN).[36] Spatially resolved 1D nXRD patterns were then further analysed in the open source Python package Hyperspy.[37]

**Nanoscale spatial maps of photoelectron spectra and work function**

Energy-resolved images for photoemission spectro-microscopy have been collected as a sequence with step of 100 meV in range from -1 to 3 eV around the valence band edge. We changed the microscope objective during the scan to maintain image focus and resolution at each energy step. Non-uniform intensity induced by dispersion in hemispherical energy analyzer has been corrected using a sample with uniform intensity as reference (gold marker). In order to determine Fermi level, we used cutoff energy at high energy side of the photoelectron spectrum taken from gold marker and plotted the spectra on binding energy scale.

To plot photoemission spectra, we extracted the photoemission intensity versus energy for each of the roughly 50 defect clusters within our 6 μm × 6 μm field of view from a stack of energy-resolved images using ImageJ software. To determine energy level position for each type of the defects, we performed multi-peak fitting using three Gaussian functions for photoemission spectra extracted from each defect cluster.

To plot work function spatial map, the PES from each pixel was extracted, and fit with a Gaussian to determine the position of the secondary edge cutoff. The work function was then calculated from $\Phi = h\omega - E_{cutoff}$ and plotted as a two-dimensional image. Energy-resolved images prior to that have been treated with flat filed correction, non-uniform intensity has been removed, and a Gaussian filter (standard deviation $\sigma = 2$) has been applied.

### Scanning electron microscopy measurements

SEM images for correlation with PEEM and photoluminescence maps have been acquired using FEI Quanta 250 FEG scanning electron microscope. The accelerating voltage was set to 5 kV with aperture of 2.5. SEM images have been correlated with PEEM images of defects using fiducial gold markers and also from comparison with PEEM images of the surface morphology as a guide. SEM measurements were performed last after all other measurements.

### Atomic force microscopy measurements

Atomic force microscopy maps have been collected in air in a tapping mode with Bruker ICON3-OS1707 atomic force microscope, and flattened with ICON3 Bruker software. The height profiles have been plotted using Gwyddion software.

### TR-PEEM trapping dynamics

Trapping dynamics for different types of defects have been extracted by recording intensity within each defect cluster of the same kind through sequence of images that corresponded to pump-probe time delays using ImageJ software. Extracted intensities have been normalized to intensity of gold fiducial marker to remove laser fluctuations. Negative time delay signal has been subtracted from positive time-delays. Subsequent extracted intensities have been averaged between multiple defects of the same kind and plotted as a function of time delay.

TR-PEEM images for each time delay have been generated by subtracting images from negative time delays from images that corresponded to positive time delays, and coloring negative change in intensity with blue, and positive with red.

The TR-PEEM curves for grain boundary defect clusters have been fitted with double-exponential equation:

$$y(t) = A_1 \times [exp(-t/\tau_1) - 1] + A_2 \times [exp(-t/\tau_2) - 1],$$

where $A_1$ and $A_2$ are amplitudes, and $\tau_1$ and $\tau_2$ are fast and slow time constants, respectively. The TR-PEEM traces for polytype defect clusters have been fitted with single-exponential equation:

$$y(t) = A_1 \times [exp(-t/\tau_1) - 1],$$

where $A_1$ and $\tau_1$ are amplitude, and fast time constant, respectively.

**Photoluminescence measurements**

Photoluminescence maps have been acquired in air using confocal microscope (Nanofinder 30). Excitation source was a 532 nm continuous wave laser. Laser probe beam was focused through the 0.8 NA 100x objective lens, and intensity of the laser was minimized to approximately 500 nW. Photoluminescence mapping was performed by scanning 10 μm by 10 μm sample area (that was prior mapped with 4.65 eV probe photons in PEEM) with the galvanic mirror with a step of 143 nm.

**Photoexposure to visible light and dry air**

Samples have been transferred to a vacuum sealed suitcase, and attached to a separate chamber with window. The chamber was pumped and purged several times, and filled with dry air. We used mild amounts of dry air and filled the chamber to approximately 1 mbar pressure. The sample was then exposed to a He-Ne laser (633 nm, less than 1 sun intensity) through the window for 1 hour; after that the sample has been transferred back to PEEM inside the vacuum sealed suitcase. During this measurement the sample was not exposed to ambient air.

**Author Contributions**

SK collected, analyzed and interpreted PEEM, TR-PEEM, confocal PL, SEM, and AFM data. AJW collected PEEM data, assisted in analysis and interpretation of data. AJW, JM and MM supported the TR-PEEM experiment, and CEP assisted in analysis and interpretation of data. NSC assisted with AFM measurements. SM prepared and provided samples. TASD, SM, KF and MA, supervised by SDS, collected, analyzed and interpreted nXRD data. SDS, SM and TASD assisted in interpretation of data. All authors contributed to discussions and manuscript preparation. KMD supervised the project.

**Conflicts of interest**

Samuel D. Stranks is a Co-Founder of Swift Solar Inc.

**Acknowledgements**

SK, AW, CEP, NSC, JM, MKLM, KMD acknowledge the support of the Femtosecond Spectroscopy Unit of the Okinawa Institute of Science and Technology, Graduate University. The authors acknowledge support for this work from the Engineering Support Section of the Okinawa Institute of Science and Technology Graduate University. KMD acknowledges support by JSPS KAKENHI Grant Number JP19KO5637. SM acknowledges funding from an Engineering and Physical Sciences Research Council (EPSRC) studentship and support from the Japan Society for the Promotion of Science (JSPS) Summer Fellowship Programme. TASD acknowledges the support of a National University of Ireland Travelling Studentship. SDS acknowledges the Royal Society and Tata Group (UF150033). The work has received funding from the European Research Council under the European Union's Horizon 2020 research and innovation programme (HYPERION - grant agreement no. 756962). The authors acknowledge the EPSRC

(EP/R023980/1) for funding. KF acknowledges a George and Lilian Schiff Studentship, Winton Studentship, the Engineering and Physical Sciences Research Council (EPSRC) studentship, Cambridge Trust Scholarship, and Robert Gardiner Scholarship. MA acknowledges funding from the Marie Skłodowska-Curie actions (grant agreement No. 841386) under the European Union's Horizon 2020 research and innovation programme. The authors acknowledge Diamond Light Source for time on Beamline I14 under Proposal sp19023, and input from Dr Paul Quinn and Dr Julia Parker.

# Supplementary Information (SI)

# Unraveling the varied nature and roles of defects in hybrid halide perovskites with time-resolved photoemission electron microscopy

Sofiia Kosar[1], Andrew J. Winchester[1], Tiarnan A. S. Doherty[2], Stuart Macpherson[2], Christopher E. Petoukhoff[1], Kyle Frohna[2], Miguel Anaya[2,3], Nicholas S. Chan[1], Julien Madéo[1], Michael K. L. Man[1], Samuel D. Stranks*[2,3], Keshav M. Dani*[1]

[1] *Femtosecond Spectroscopy Unit, Okinawa Institute of Science and Technology, 1919-1 Tancha, Onna-son, Okinawa, 904-0495, Japan.*

[2] *Cavendish Laboratory, University of Cambridge, JJ Thomson Avenue, Cambridge CB3 0HE, United Kingdom.*

[3] *Department of Chemical Engineering and Biotechnology, University of Cambridge, Philippa Fawcett Drive, Cambridge CB3 0AS, United Kingdom.*

*kmdani@oist.jp, *sds65@cam.ac.uk

Fig. S1 Photoluminescence spectroscopy for perovskite thin film sample.

Fig. S2 Spatially averaged x-ray diffraction pattern.

Fig. S3 Energy-resolved PEEM and size distribution of defect clusters.

Fig. S4 Additional photoemission spectra and work function maps.

Fig. S5 Additional TR-PEEM images and traces.

Fig. S6 Additional regions with photoluminescence (PL) maps around defect clusters (PEEM), and SEM images for selected defects.

Fig. S7 Additional photoemission spectra and work function maps after treatment with dry air and light.

Fig. S8 Response of polytype and PbI2 defect clusters upon exposure to light and dry air.

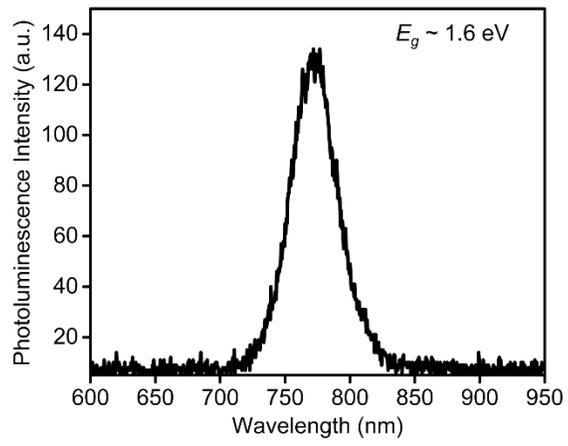

**Fig. S1 Photoluminescence spectroscopy for perovskite thin film sample.** PL spectrum with 532 nm excitation.

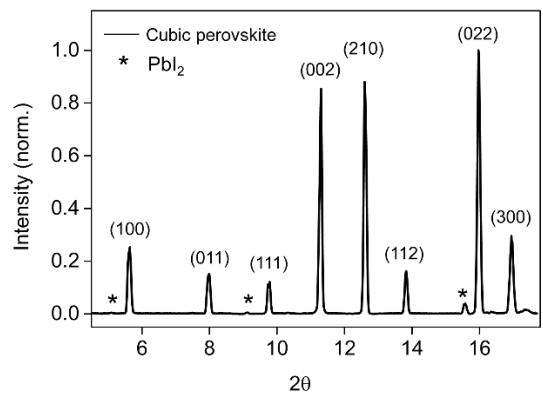

**Fig. S2 Spatially averaged x-ray diffraction pattern**. Extracted from the same region as shown in Fig. 1c, indexed to a cubic perovskite structure, with detected $PbI_2$ peaks.

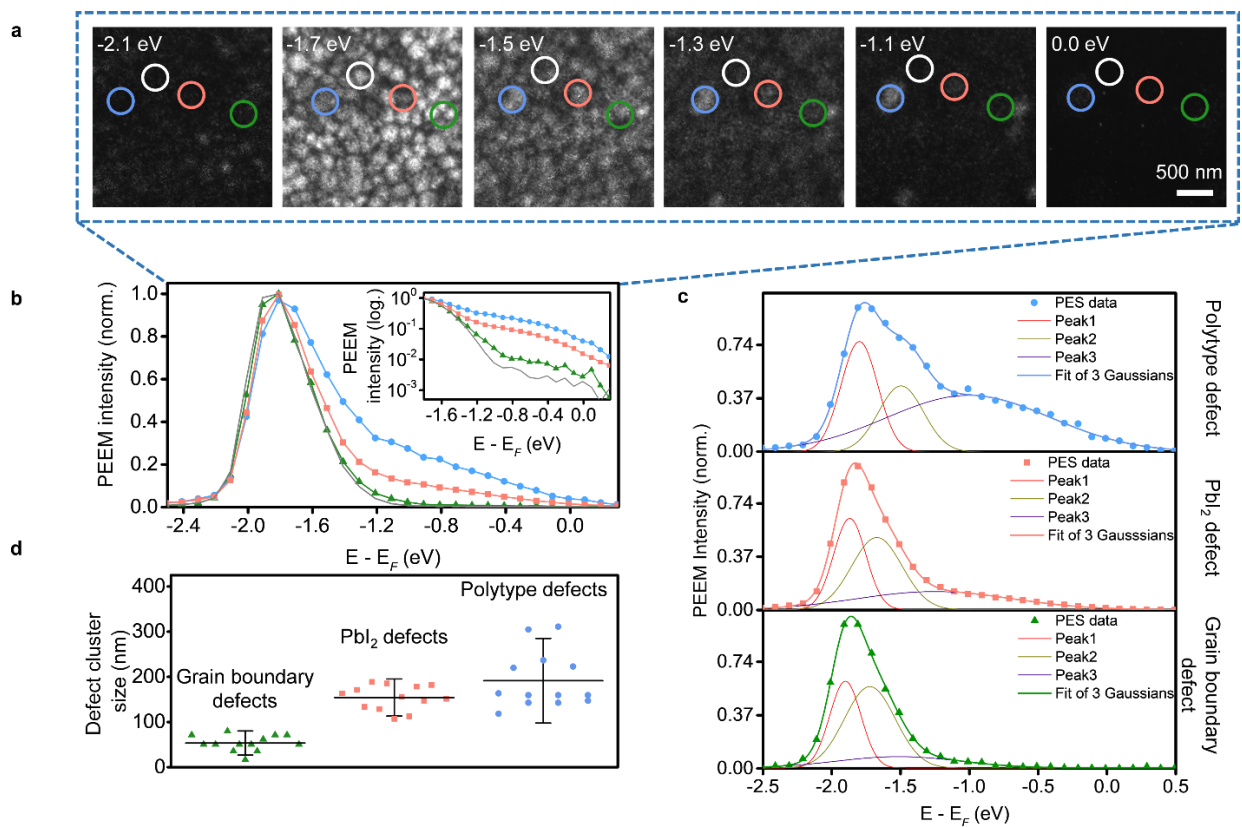

**Fig. S3 Energy-resolved PEEM and size distribution of defect clusters. a** Representative set of energy-resolved PEEM images showing surface morphology at (~ - 1.7 eV) and appearance of localized defect states as clusters at lower energies. **b** Photoelectron spectra for three types of defect clusters, averaged for multiple clusters of the same type, and region without defects (grey solid line). **c** Gaussian fits (red, yellow, purple) to the photoelectron spectra for selected defect clusters used to estimate peak energy of the mid-gap defect states for three types of defect clusters. **d** Size distribution for three types of defect clusters, as labelled; solid black line indicates the mean value, error bars show standard deviation.

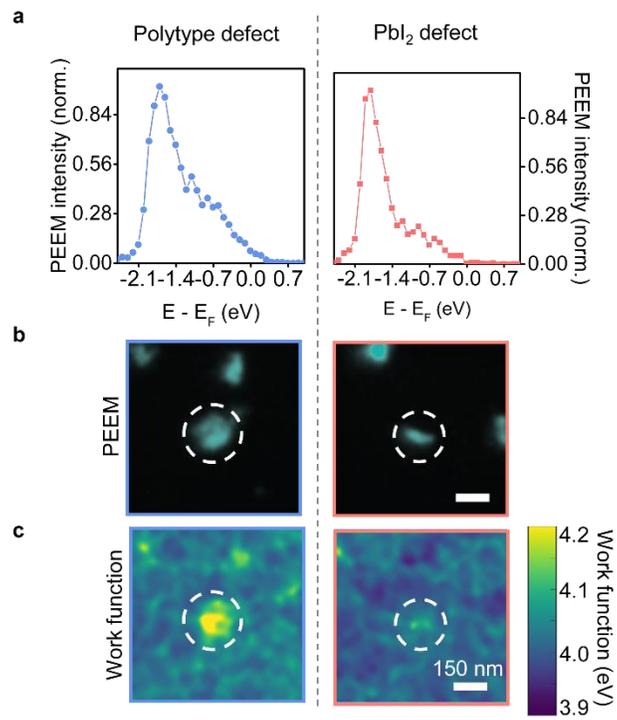

**Fig. S4 Additional photoemission spectra and work function maps**. **a** Photoemission spectra for polytype and PbI$_2$ defect clusters shown in PEEM images in **b**. **c** Work function maps with nanoscale resolution for the region around the polytype and PbI$_2$ defect clusters shown in (b).

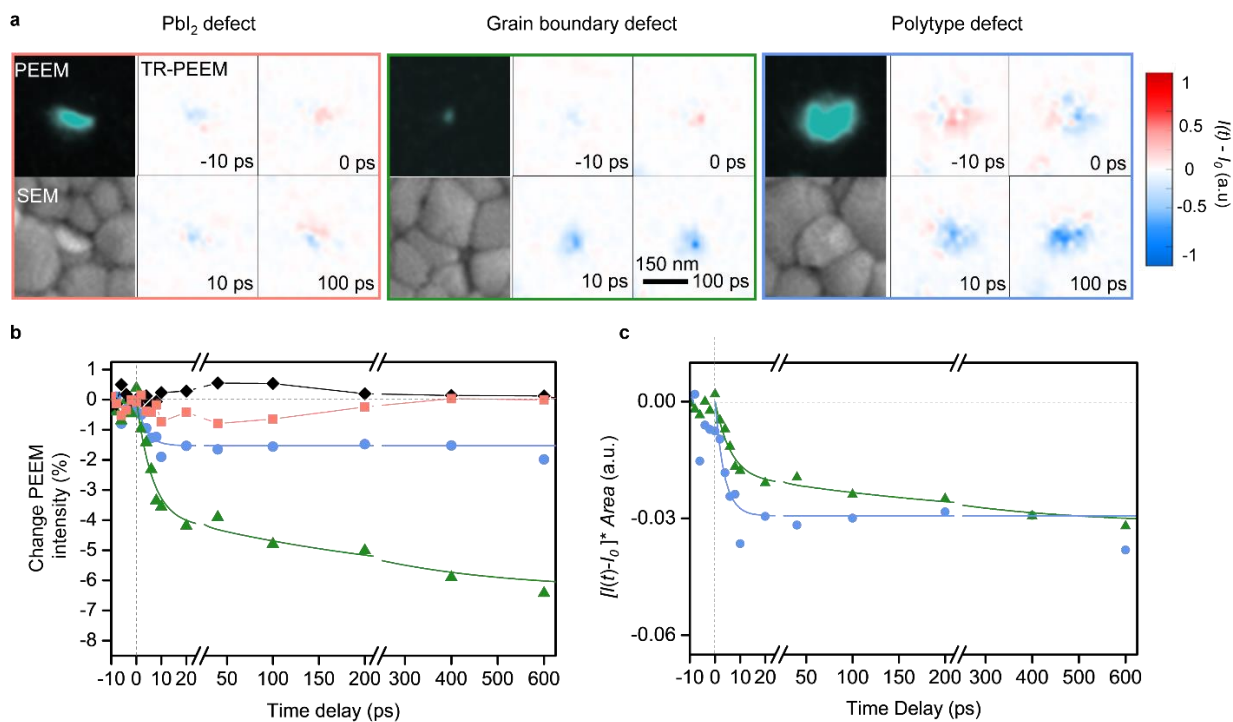

**Fig. S5 Additional TR-PEEM images and traces**. **a** PEEM and SEM images of three types of defect clusters with corresponding TR-PEEM intensity changes for selected pump-probe delays. $I_0$ indicates PEEM intensity before the photoexcitation, $I(t)$ corresponds to PEEM intensity for each time delay after the photoexcitation. **b** TR-PEEM traces indicating percentage change in PEEM intensity plotted as $[I(t) - I_0]/I_0 \times 100$ for three types of defects and region without defects. Solid lines represent bi-exponential fit for grain boundary defects (green) with time constants $\tau_1$ = 6.4 ps ± 1.5 ps, $\tau_2$ = 291 ps ± 162 ps and amplitudes $A_1$ = 4.03 ± 0.42, $A_2$ = 2.28 ± 0.41; and single exponential fit for polytype defects (blue) with time constant $\tau_1$ = 3.7 ps ± 1.04 ps, and amplitude $A_1$ = 1.53 ± 0.08. **c** TR-PEEM traces for polytype and grain boundary defect clusters plotted as intensity change $[I(t)-I_0] \times Area$, solid lines represent single and double exponential fits.

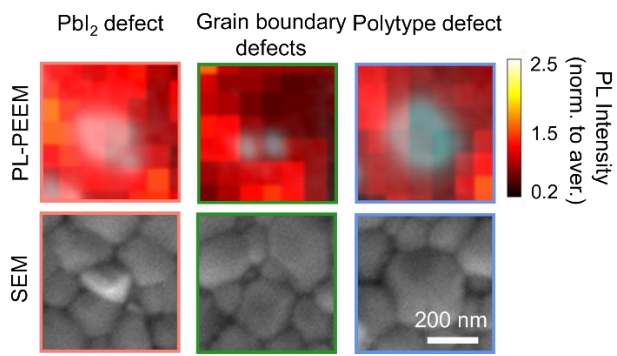

**Fig. S6 Additional regions with photoluminescence (PL) maps around defect clusters (PEEM), and SEM images for selected defects.** Top row shows PEEM images (green blur) overlaid on PL maps for three types of defect clusters, with corresponding SEM images (bottom row).

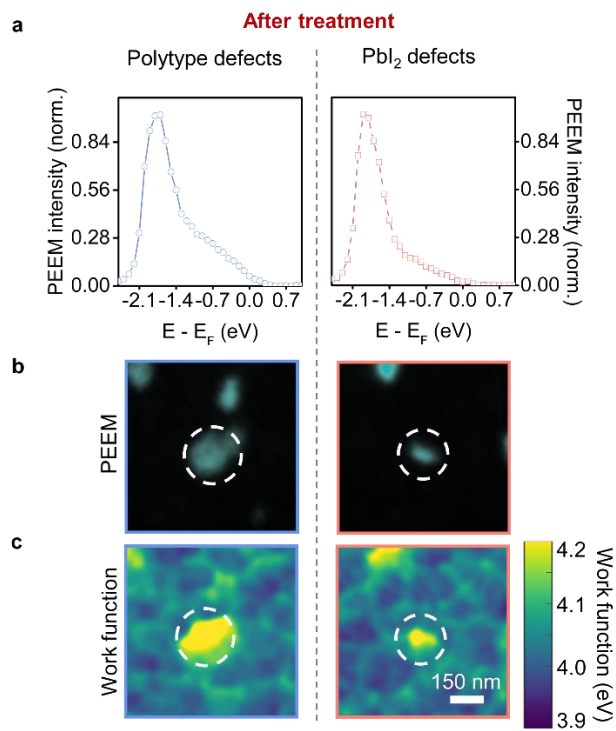

**Fig. S7 Additional photoemission spectra and work function maps after treatment with dry air and light**. **a** Photoemission spectra for polytype and PbI$_2$ defect clusters shown in PEEM images in **b**, with nanoscale work function maps in **c**. While the density of mid-gap states is not largely affected after treatment with dry air and light, the work function increased for sites of PbI$_2$ defect clusters. Results shown for the same defect clusters as in Fig. S4.

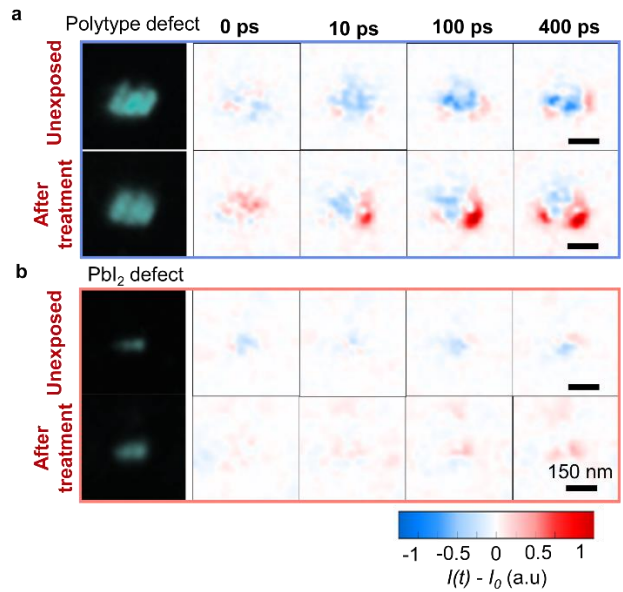

**Fig. S8 Response of polytype and PbI$_2$ defect clusters upon exposure to light and dry air. a** PEEM and TR-PEEM images before and after the exposure to light and dry air for polytype defect clusters. **b** PEEM and TR-PEEM images before and after the exposure to light and dry air for PbI$_2$ defrct clusters.